\documentclass[final,5p,times,twocolumn]{elsarticle}
\usepackage{amssymb}
\usepackage{graphicx}
\usepackage{dcolumn}
\usepackage{amsmath}

\begin{document}

\begin{frontmatter}

\title{
Topological nature of Hubbard bands in strongly correlated systems}

\author{V. Yu. Irkhin}

\ead{Valentin.Irkhin@imp.uran.ru}
\author{Yu. N. Skryabin}
\address
{M. N. Mikheev Institute of Metal Physics, 620108 Ekaterinburg, Russia
}

\begin{abstract}
  The topological nature of the Mott-Hubbard state in strongly correlated systems is treated. These systems are described in terms of spin-charge separation, i.e. spinon-holon deconfinement in the gauge field.
  Analogies with the quantum Hall effect and Landau quantization are considered,  the presence of the Chern-Simons term being important for  proper description of the Hubbard splitting.  Occurrence of topological frustrations with doping in an effective two-band model is demonstrated.
  The role of orbital degrees of freedom and possible occurrence of orbital currents are discussed.
\end{abstract}



\end{frontmatter}
%

Last years, the topological description of Mott-Hubbard  systems has been developed, including  unusual superconductivity and algebraic spin liquid phase of the gauge theory \cite{Wen,Hermele,Scr3}.
At the same time, traditional methods of theoretical consideration of strongly correlated electron systems (e.g., diagram technique, equation of motion decoupling etc.) are usually not related to the quantum-topological picture. 

In the present Letter we try to provide such a relation within slave-particle representations.
In particular, we relate the topological concepts to the standard description of the correlation Hubbard splitting which  is a characteristic feature of the systems with strong correlations  \cite{Hubbard,Hubbard2}.
We demonstrate the analogy of the Hubbard subbands with Landau levels in the gauge field in the situation of the quantum Hall effect.
We treat also the problem of orbital currents in the doped systems which has been recently  investigated by numerical methods \cite{Weng1}.

We start from the Hamiltonian of the $t-J$ model
\begin{equation}
\mathcal{H}=\sum_{ij\sigma }t_{ij}X_i(0,\sigma )X_j(\sigma,0)
+\mathcal{H}_{d}.
 \label{t-J}
\end{equation}
with $\mathcal{H}_{d}=\sum_{ij}J_{ij}{\bf S}_i {\bf S}_j$ the Heisenberg Hamiltonian,
\begin{equation}
X_i(0,\sigma ) = |i0 \rangle \langle i\sigma|={c}_{i\sigma } (1-{c}^\dag_{i-\sigma }{c}_{i-\sigma })
\end{equation}
are the projection Hubbard operators, $|i0 \rangle$ and $|i\sigma\rangle$ are empty  and singly-occupied on-site states.

The spin-liquid state is treated in terms of exotic quasiparticles -- spinons and holons. These were introduced by Anderson \cite{633a} to describe two-dimensional cuprates as
\begin{equation}
\tilde{c}_{i\sigma }=X_i(0,\sigma )=b_i^{\dagger }f_{i\sigma }.
\label{eq:6.131c}
\end{equation}
According to Anderson, spinons  $f_{i\sigma}$ obey the Fermi statistics   and holons $b_i$ the Bose statistics.  This choice is not unique and can be changed depending on the physical problem, e.g., in the presence or absence of magnetic ordering, see Ref.\cite{Kane}.
When using the representation (\ref{eq:6.131c}) in simple mean-field treatments, the problem of non-physical states occurs: the on-site no-double-occupancy constraint is violated.
Thus spinons and holons become coupled by a gauge field
required to satisfy this constraint.

The Mott metal-insulator transition is described by the condensation
of charged  bosonic holons $b$. A more general SU(2) representation \cite{Wen} introduces two kinds of bosons,
\begin{align}
\tilde{c}_{i\uparrow }& ={\frac{1}{\sqrt{2}}}\left( b_{i1}^{\dagger
}f_{i\uparrow }+b_{i2}^{\dagger }f_{i\downarrow }^{\dagger }\right),  \nonumber
 \\
\tilde{c}_{i\downarrow }& ={\frac{1}{\sqrt{2}}}\left( b_{i1}^{\dagger
}f_{i\downarrow }-b_{i2}^{\dagger }f_{i\uparrow }^{\dagger }\right).
\label{22}
\end{align}

Fluctuations of the gauge field are essentially chirality fluctuations or fluctuations of orbital current. The corresponding staggered flux  (SF) phases are obtained in the slave-boson mean-field approach \cite{Wen,Wen1}. The SF state is competing with d-wave superconductivity and antiferromagnetic (AFM) ordering in systems with nodal points (high-$T_c$ cuprates).

A distinctive feature of the Gutzwiller projected SF state is that it breaks translational symmetry and orbital currents circulate the plaquettes at finite doping.
Projected SF  and d-wave superconductor states  are closely related owing to the SU(2) symmetry.
The latter state does not break translational or time-reversal symmetry, and possesses no static current, but demonstrates a power-law correlation
in AFM order and staggered orbital current.
At finite doping the projection of a SF phase  possesses long-range orbital current order, but short-range pairing and AFM
order.
The projected SF state is less favorable than  the  d-wave superconductor, but the energy difference is small and vanishes as doping goes to zero.



A way to obtain a stable deconfined phase is breaking the U(1) or
SU(2) gauge structure down to a  Z$_2$ gauge structure \cite{Wen}.
Such a phase is called a Z$_2$ spin liquid or a short-range RVB state.
The Z$_2$ spin state also contains neutral spin-1/2 excitations.
The SF state  is an example of collinear SU(2) flux state which is invariant only under a U(1) rotation. This is a marginal situation.

The state with noncollinear SU(2) flux is a Z$_2$ state. In a Z$_2$ state, all the gauge fluctuations are gapped. Here the fluctuations can only mediate short-range interactions between fermions.  Therefore including fluctuations does not qualitatively change the properties of the mean-field state, the gauge interactions are irrelevant and the Z$_2$ mean-field state is stable at low energies. This means  the existence of a real physical spin liquid which contains fractionalized spin-1/2 neutral fermionic excitations. This spin liquid also contains a Z$_2$ vortex excitation, so-called visons. The bound state of a spin-1/2 fermionic excitation and a Z$_2$ vortex gives us a spin-1/2 bosonic excitation.

Another way to obtain a deconfined phase is to give the gauge boson a mass.
Most simple description of such a situation is given in terms of the chiral spin liquid where the Chern-Simons term is present.  The picture is as follows  \cite{Wen1}.
The  excitation in the mean field approximation are obtained by adding a spinon to the conduction band. However, this excitation is not yet physical, since the spinon in the conduction band violates the constraint $\sum_{\sigma} \langle f^\dag_{i\sigma}f_{i\sigma}\rangle=1$.
The additional density of spinons can be eliminated by introducing a vortex flow of the gauge field
\begin{equation}
\Phi= -\pi \sum_{i} \left(\sum_{\sigma}\langle f^\dag_{i\sigma}f^{}_{i\sigma}\rangle-1\right).
\label{fi}
\end{equation}
Therefore the physical quasiparticles are spinons dressed by a $ \pi $ vortex which carry spin 1/2. At the same time, spinon, which carries a single charge of the gauge field, has fractional (semion) statistics, being a bound  state of charge and vortex \cite{Wen1}. After turning off the lattice potential, the valence band in the mean-field chiral spin state becomes the first Landau level, so that the ``Landau levels'' arising in the ``electromagnetic'' gauge field correspond to the Hubbard bands.
Thus we have orbital quantization in an intrinsic gauge field, which determines correlated band structure with narrowed bands.


A similar treatment can be  developed within the doubled Chern-Simons theories
where macroscopic chirality and time reversal violation is absent.
Such a theory was considered by Levin and Wen \cite{Levin} who investigated  mutual statistics of spinon and vison excitations which can be treated as mutual semions.
Here, adiabatically moving a spinon around a vison  yields a Berry phase of $ \pi$. Thus the charges move on the square lattice, while the fluxes  on the dual lattice. Formulation of some three-dimensional models is also possible \cite{Levin}.

Mutual semion statistics of spinons and holons in superconducting and AFM phases was also considered in the $t-J$ model  within slave-fermion approach of Weng's phase string theory with application to cuprate systems \cite{Weng}.
According to this theory,
in the underdoped regime the AFM and superconducting phases
are dual: in the former, holons are confined while spinons are deconfined,
and vice versa, and the gauge field, radiated by the holons (spinons),
interacts with spinons (holons) through minimal coupling.
The  corresponding  mechanism of Luttinger-liquid-like behavior is  Anderson's  unrenormalizable Fermi-surface phase shift generated by the doped holes; this was identified  with a many-body Berry-like phase \cite{Weng2,Weng1}.
Thus Weng's theory introduces topological defects with nontrivial quantum numbers, which correspond to vortices. Weng's Lagrangian possesses both parity and time-reversal symmetry; a similar structure for Z$_2$ spin liquid in terms of spinons and visons was obtained in Ref.~\cite{Sachdev2}.

For an arbitrary loop $C$ on the lattice  one gets \cite{Weng}
\begin{eqnarray}
\sum_{\langle i,j \rangle\in C}\, A_{ij}^s &  =  & \pi \sum_{l\in
S_C}\, \left( b^\dagger_{l\uparrow}
b^{}_{l\uparrow}-b^\dagger_{l\downarrow} b^{}_{l\downarrow}\right) \,\, {\rm mod}\, 2\pi\,,\nonumber\\
\sum_{\langle i,j \rangle\in C}\, A_{ij}^h &  =  & \pi \sum_{l\in
S_C}\, h^\dagger_l h^{}_l \,\, {\rm mod}\, 2\pi \,, \label{66}
\end{eqnarray}
where $b$ and $h$ are Bose spinon and holon operators, $A_{ij}^{s,h}$ are the gauge fields, the sum on the left-hand side runs over all the links on
$C$, and on the right-hand side over all the sites
inside $C$. Thus the holon (spinon) carries a $\pi$-flux and couples to the motion of spinons (holons) via  $A_{ij}^h$ ($A_{ij}^s$).
Thereby the quantization of the flux of the gauge fields with unit value of $\pi$ results in integer values of spinon and holon occupation numbers.
 As well as the consideration of vortices in the chiral spin liquid \cite{Wen1}, mutual Chern-Simons term in \cite{Weng} enables one to realize the no-double-occupancy constraint, i.e. to describe properly Hubbard's projection.



New understanding was achieved owing to discovery of topological insulators \cite{Hasan}. Conventional topological insulators (with bulk energy  gap and  gapless edge states) are similar to the integer quantum Hall effect states. Here, owing to the presence of a single-particle energy gap, electron-electron interactions do not modify the state in an essential way. Thus they can be understood within the framework of the band theory of solids. Therefore the simpler integer quantum Hall states are adequately described in terms of single-particle quantum mechanics (however, they can be treated in terms of the so-called invertible topological order \cite{Wen3}).
Recent work \cite{Rubtsov} treats correlation effects at finite temperatures in the quantized integer-valued Hall conductivity  corresponding to a topological invariant, the first Chern number.
Since the Hall conductivity is odd under time reversal the topologically nontrivial states described  can only occur when this symmetry is broken. However, the spin-orbit interaction allows a different Z$_2$ topological class
of insulating band structures when the symmetry is unbroken \cite{Hasan}.

On the contrary, the  correlated fractional quantum Hall states
are characterized by topological order and quantum entanglement, and require an essentially many-body treatment.
The Hubbard systems where electron correlations play the crucial role are similar to the latter states.


Excitations of the Dirac spin liquid considered in \cite{Vishwanath}  include magnetic monopoles that drive confinement. When the fermions fill a Chern band, a Chern-Simons term is generated that represents charge-flux attachment,
similar to spin being carried by the monopole (a flux quantum) in the presence of a quantized spin Hall conductance.
Stability of the spin liquid state is enhanced on the frustrated (e.g., triangular and kagome) lattices as compared to the bipartite (square and honeycomb graphene) lattices.

\begin{figure}[h]
  \includegraphics[width=0.45\textwidth]{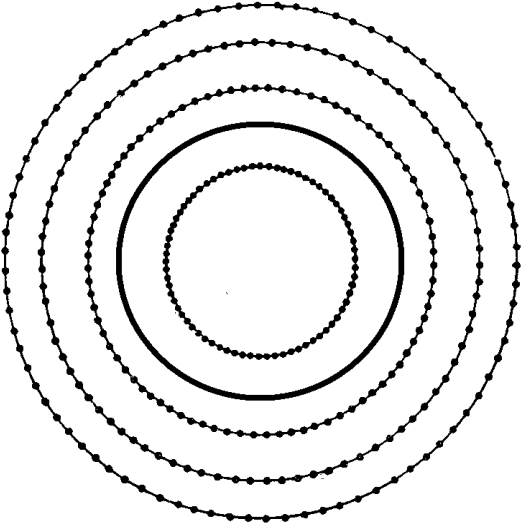}
  \caption{Landau quantization in the gauge field in the $k_x-k_y$ plane. The points denote density of electron states on the levels. The bold line shows the Fermi energy in the insulating state.}
\label{2}
\end{figure}

Proliferation of monopoles results in boson condensation and occurrence of gaps in the Dirac points. In the presence of fluxes, quantization  in the gauge field results in occurrence of Landau-type levels, see Fig. 1.
In particular, a zero energy Landau level has a degeneracy equal to the number of flux quanta $\Phi/2 \pi$. Addition of a flux quantum creates a fermion zero mode for each Dirac fermion flavor.
After inclusion of crystal lattice potential, they become transformed into narrow correlated bands. In this sense, the Hubbard bands (which are described in the simplest Hubbard-I approximation \cite{Hubbard} as broadened atomic levels) are spinon bands.
Similar bands are obtained in the rotor representation
\begin{equation}
c_{i\sigma }^{\dag }=f_{i\sigma }^{\dag }\exp (i\theta _{i})
\end{equation}
where  the original Hubbard interaction Hamiltonian is written in terms of the the angular momentum $\hat{L}=-i\partial /\partial \theta $ as
 $(U/2)\hat{L}_{i}^{2}$ \cite{rotor1}, the spinon spectrum being renormalized by $Z$-factors. This consideration is generalized to the case of orbital-degenerate bands, as well as the Hubbard's consideration \cite{Hubbard2}.
Again, the situation is analogous to the conventional Landau quantization in magnetic field, where the energy levels become almost $\bf k$-independent \cite{ziman}. However, in our case we deal with the quantization for spinons rather than for free electrons, see Eq.(\ref{fi}).

The non-Landau-symmetry-breaking AFM Mott insulator--superconductor transition was considered in Ref. \cite{Vishwanath1} within a special model in the rotor SU(2) representation,  the ordered phases being intimately related to a topological band insulator. The transition is characterized by nontrivial vortex quantum numbers: the vortices of the antiferromagnet are charged and the vortices of the superconductor carry spins.

The topological interaction can be  treated within a field theory (see. e.g.. Ref. \cite{Oganesyan}).
In the 2+1 dimensional case  both quasiparticles
and vortices are particles, so that  one has a close analogy to
the bosonic Chern-Simons theory for the quantum Hall effect and we can attach flux and charge to the particles in such a way that the Berry phases
appear.  The situation is similar to the  Aharonov-Bohm effect, which can be accompanied by the fermion--boson statistics transmutation.
The wavefunction acquires a well defined Berry phase given by the line integral of the Berry connection. This may be expressed as a surface integral of the Berry curvature. The Chern invariant is the total Berry flux  in the Brillouin zone \cite{Hasan}.

A unit charge quasiparticle current $j^{\mu}$, and a vortex current $J^{\mu}$ are coupled  to electric and magnetic gauge potential components
$a_{\mu}$ and $b_{\mu}$ via the Chern-Simons Lagrangian \cite{Oganesyan},
\begin{equation}
{\cal L} =
\frac 1 \pi \epsilon^{\mu\nu\sigma}b_\mu\partial_\nu a_\sigma
 - a_{\mu}j^{\mu} - b_{\mu}J^{\mu} \, .
\end{equation}
In 3+1 dimensions,
the vortices become strings, and the vector potential $b$ is  an
antisymmetric tensor $b_{\mu\nu}$.

The above approaches treat essentially an insulating case. It should be noted that the Mott transition can occur at finite doping \cite{Wen} (the insulator state at finite doping can be described also in the rotor representation \cite{rotor}).

According to the consideration in Refs. \cite{Bulaevskii,Motrunich}, orbital current contributions occur in higher orders of perturbation theory even in the the half-filled insulating situation.
The old ideas of exciton condensation \cite{Vons} and toroidal ordering in crystals \cite{Tugushev} should be also mentioned in this context.

However, more strong effects can be induced by frustrations owing to motion of  current carriers \cite{Wen}.
These effects are described by the scalar spin chirality $\chi_{ijk}=({\bf S}_i [{\bf S}_j\times{\bf S}_k])$  (note that the SU(2) state \cite{Wen}, Eq.(\ref{22}), is not characterized by chirality since another order parameter is used).
Thus, in the doped case, two channels should occur which are connected with spinons (fractionalized degrees of freedom) and conduction electron states.


A microscopic description can be obtained within the dopon representation \cite{Ribeiro}
which was used in the problem of nodal-antinodal dichotomy in high-$T_c$ cuprates \cite{Wen}. The corresponding expression for the Hubbard operator reads
\begin{equation}
X{_i(0, -\sigma) }=-\frac \sigma{2\sqrt{2}}\sum_{\sigma ^{\prime }}d_{i\sigma
^{\prime }}^{\dagger }(1-n_{i,-\sigma ^{\prime }})
[\delta _{\sigma \sigma ^{\prime }}-2(\mathbf{S}_i\mbox{\boldmath$\sigma $}_{\sigma ^{\prime }\sigma} )].
 \label{eq:I.8}
\end{equation}
where the Fermi dopon operators $d_{i\sigma^{\prime }}^{\dagger }$ describe current carriers, and spin operators $\mathbf{S}_i$ the localized degrees of freedom; they can be represented in terms of Fermi or Bose (Schwinger)   spinons \cite{Ribeiro,Punk,Skryabin}.
Density of doped charge carriers $x$ equals to the density of dopons,
\begin{equation}
x=1-n=\sum_{\sigma}\langle n_{i,\sigma}  \rangle,\,  n_{i,\sigma} =d^\dag_{i\sigma}d^{}_{i\sigma},
\label{eq:I.71}
\end{equation}
which is an exact relationship valid in the projected subspace.
On substituting (\ref{eq:I.8}) into the $t-J$ Hamiltonian (\ref{t-J})
we have
\begin{multline}
\mathcal{H}=\frac{1}{(2S+1)^{2}}\sum_{ij\sigma \sigma ^{\prime }}t_{ij}\{(S^{2}+%
\mathbf{S}_{i}\mathbf{S}_{j})\delta _{\sigma \sigma ^{\prime }}-S(\mathbf{S}%
_{i}+\mathbf{S}_{j})\mbox{\boldmath$\sigma $}_{\sigma \sigma ^{\prime }} \\
+i\mbox{\boldmath$\sigma $}_{\sigma \sigma ^{\prime }}[\mathbf{S}_{i}\times
\mathbf{S}_{j}]\}d_{i\sigma }^{\dagger }(1-n_{i,-\sigma })(1-n_{j,-\sigma
^{\prime }})d_{j\sigma ^{\prime }}+\mathcal{H}_{d}  \label{eq:I.10}
\end{multline}
with $S=1/2$. Such a representation of the Hamiltonian in the narrow-band $s-d$ model with arbitrary $S$ (being also valid in its particular case -- the $t-J$ model) was obtained  in \cite{II}. 

Thus we obtain the terms  which are linear in spin operators. These can provide hybridization between electrons (dopons) and Fermi spinons to describe the dichotomy.

Therefore the initial one-band model takes the form of an effective two-band model,  as well as for the Kondo lattices problem.
Here, orbital-selective (partial) Mott transition in one band takes place which is a quantized change of Fermi surface, i.e. transition from large to small Fermi surface  \cite{Vojta}, which is related to formation of the Hubbard subbands.

Spinon-dopon pairing  in the dopon approach \cite{Ribeiro} leads to the same phases as the holon condensation in the slave-boson formalism \cite{Wen}. In this sense, superconductivity is condensation of auxiliary boson or spinon-dopon pair.

Although calculations with the Hamiltonian (\ref{eq:I.10})  in the absence of a small parameter are difficult, it provides a natural description of chirality and orbital current owing to vector product terms.

According to numerical results using exact diagonalization and density matrix renormalization group methods \cite{Weng1}, a nontrivial many-body Berry-like phase and persistent spin currents are indeed revealed in the ground state of the $t-J$ model with periodic and  open boundary condition. 
They are accompanied by a nonzero total momentum or angular momentum
associated with the doped hole. This determines a nontrivial ground state degeneracy.






A general topological analysis of the Luttinger theorem (which determines the Fermi surface volume) was performed by  Oshikawa  \cite{Oshikawa,Sachdev}. This is is based on momentum balance argument,  global gauge symmetry U(1) (charge conservation) and threading a $2\pi$ quantum of flux.
Due to cyclic boundary conditions, the system is considered as a torus, in the contours of which a crystal momentum arises; this is similar to the appearance of Faraday force with flux variation.
The  change of the crystal momentum corresponding to the flux insertion is determined by the number of momentum space points inside the Fermi surface.
Each ``Landau  energy'' level  has the degeneracy $\Phi/(2 \pi)$ in the units of flux quanta. This degeneracy equals to the number of states between two quantized orbits
\cite{ziman}, in agreement with Eq.(\ref{fi}).


Thus, if we take into account global topological excitations, the violation of the Luttinger theorem, similar to fractional Hall effect state, should be accompanied by the topological order  in the spin-liquid state.
 The corresponding exotic non-magnetic fractionalized Fermi-liquid state FL$^*$ in the two-band model possesses small Fermi surface  and can be of U(1) or Z$_2$ type \cite{Sachdev}.
The low-energy excitations of the state on a torus are given by the action
${\cal S}_{\rm FL^*} = {\cal S}_{\rm FL} + {\cal S}_{\rm CS}$, which is the direct sum of the action for fermionic quasiparticles, and of the topological Chern-Simons action \cite{Sachdev1}.

 The related phenomena are  the formation of  a paramagnetic state with quantum-disordered (frustrated) local moments (which also means a  non-Fermi-liquid behavior) and  Mott-Hubbard splitting. Indeed, the Oshikawa theorem holds for both  insulator and metal (and even in the 3+1 dimensional  case \cite{Sachdev}), and the Mott transition is a
quantized change in the Fermi volume.

The FL$^*$--FL transition with increasing doping is most conveniently described in terms of fermionic spinons. On the other hand,  AFM--FL$^*$ transition with increasing frustration uses bosonic spinons (e.g., Weng's theory \cite{Weng} or Schwinger dopon approach \cite{Punk}), a description of the transmutation of the neutral spinon excitations of the FL$^*$ phase from fermions to bosons being still absent \cite{Punk1}.

The deconfinement gauge theory including spin-charge separation and analogies with the quantum Hall effect was historically formulated in the 2+1 case. At the same time, some generalizations to the 3+1 case are also possible; this problem needs further investigations. In particular, one can construct bosonic models on a cubic lattice that have emergent gapless photons (U(1) gauge bosons), see \cite{Wen} and references therein. The Mott transition between a spin-liquid insulator and a metal can be also treated in three dimensions using the slave rotor approach \cite{Podolsky}. For the corresponding treatment of 3+1 topological insulators, see Ref.\cite{Armitage}.


To conclude, topological effects and gauge field are closely related to the correlation Hubbard splitting.
Frustrations created by current carriers can result in formation of exotic states with topological order and orbital degrees of freedom.
The   effective two-band model including spinons and conduction states provides a corresponding description.

The research was carried out within the state assignment of FASO of Russia (theme ``Flux'' No AAAA-A18-118020190112-8 and theme ``Quantum'' No. AAAA-A18-118020190095-4) and supported in part by the RFFI grant No. 17-52-52008.

%
%
%
%
%
%
%
%

{}

\end{document}